\documentclass[10pt,aps,pra,twocolumn]{revtex4-1}
\usepackage{graphicx}
\usepackage{amssymb}
\usepackage{amsmath}
\usepackage{appendix}
\usepackage{comment}

\def\p{\partial}
\def\i{\imath}
\def\j{\jmath}

\def\bb{\mathbf{b}}

\def\bs{\mathbf{s}}

\def\bv{\mathbf{v}}
\def\bw{\mathbf{w}}
\def\bz{\mathbf{z}}

\def\bH{\mathbf{H}}

\def\bS{\mathbf{S}}

\def\fa{\mathfrak{a}}
\def\fS{\mathfrak{S}}
\def\FH{\hat{\mathfrak{H}}}
\newcommand{\rf}[1]{(\ref{#1})}
\newcommand{\al}[1]{\begin{aligned}#1\end{aligned}}
\newcommand{\ar}[2]{\begin{array}{#1}#2\end{array}}
\newcommand{\eq}[1]{\begin{equation}#1\end{equation}}

\usepackage{mdframed}

\begin{document}

\title{Exceptional points in dissipatively coupled spin dynamics}

\author{Yaroslav Tserkovnyak}
\affiliation{Department of Physics and Astronomy, University of California, Los Angeles, California 90095, USA}

\begin{abstract}
We theoretically investigate dynamics of classical spins exchange-coupled through an isotropic medium. The coupling is treated at the adiabatic level of the medium's response, which mediates a first-order in frequency dissipative interaction along with an instantaneous Heisenberg exchange. The resultant damped spin precession yields exceptional points (EPs) in the coupled spin dynamics, which should be experimentally accessible with the existing magnetic heterostructures. In particular, we show that an EP is naturally approached in an antiferromagnetic dimer by controlling local damping, while the same is achieved by tuning the dissipative coupling between spins in the ferromagnetic case. Extending our treatment to one-dimensional spin chains, we show how EPs can emerge within the magnonic Brillouin zone by tuning the dissipative properties. The critical point, at which an EP pair emerges out of the Brillouin zone center, realizes a gapless Weyl point in the magnon spectrum. Tuning damping beyond this critical point produces synchronization (level attraction) of magnon modes over a finite range of momenta, both in ferro- and antiferromagnetic cases. We thus establish that damped magnons can generically yield singular points in their band structure, close to which their kinematic properties, such as group velocity, become extremely sensitive to the control parameters.
\end{abstract}

\maketitle

\section{Introduction}

Over the past several decades, it became clear that a class of unconventional degeneracies are abundant in diverse physical systems undergoing non-Hermitian dynamics \cite{heissJPAT12}. This generally concerns a Schr{\"o}dinger-type evolution of a complex-valued vector field $\bv(t)$:
\eq{
i\frac{d}{dt}\bv=H\bv\,,
\label{SE}}
in terms of a non-Hermitian matrix-valued ``Hamiltonian" $H(\bw)$, parametrized by a set of complex-valued system parameters $\bw$. The dynamics governed by Eq.~\rf{SE} become extremely sensitive to the values of $\bw$ close to points $\bw_0$ where $H$ is not diagonalizable. In fact, the standard diagonalization of $H$ at $\bw_0$ would yield a branch point singularity dubbed exceptional point (EP) \cite{katoBOOK66}. The defining property of $\bw_0$ is the degeneration of two or more eigenvectors at an eigenfrequency degeneracy point. For example, a spin-$1/2$ raising operator $\hat{\sigma}_+=\hat{\sigma}_x+i\hat{\sigma}_y$, expressed in terms of two $2\times2$ Pauli matrices $\hat{\sigma}_i$, has a sole eigenvector, with zero eigenvalue, corresponding to the spin-up state. As such, $\hat{\sigma}_+$ cannot be diagonalized by a similarity transformation.

The eigenenergy solutions $\lambda$ of $H(\bw)$ realize, as a function of $\bw$, multivalued functions with branch cuts terminating at branch points. Smoothly varying one of the complex-valued parameters $\bw$, the eigenenergies can be represented as single-valued functions on a multisheet Riemann surface, with branch points marking coalescence of two or more energy levels. These EPs thus provide genuine singularities, which can manifest prominently in microwave \cite{dembowskiPRL01,*dopplerNAT16} and optical \cite{hodaeiNAT17,*miriSCI19} response properties and hybrid dynamical systems \cite{xuNAT16,zhangNATC17,bernierPRA18,harderPRL18}, scattering problems and sensing \cite{chenNAT17}, open and quasi-Hermitian quantum systems \cite{benderRPP07}, etc. \cite{heissJPAT12}, particularly in regard to their topological encircling aspects \cite{berryANY95,*mailybaevPRA05,*zhongNATC18,*zhangPRX18}. The emergence of EPs in quasiparticle band structures, furthermore, tremendously enriches their topological classification in crystalline materials \cite{kawabataPRX19}.

In this paper, we argue that EPs are also commonplace in pure spin dynamics, based on several generic examples, even in the absence of external driving, such as spin-transfer torque \cite{galdaPRB16,*galdaPRB18,*galdaCM19}. In particular, an isotropic antiferromagnetic spin pair harbors an EP already in its singlet-like ground state. We show how this EP gets inherited by extended antiferromagnetic dynamics, manifesting in Weyl singularities and synchronization (level attraction) within magnonic band structure, which are tunable by the dissipative properties of the environment. While somewhat less natural, similar EPs can also be engendered by ferromagnetic systems.

The paper is structured as follows: In Sec.~\ref{gtsd}, a general model for dissipatively coupled spin dynamics is formulated, based on exchange and spin-pumping mediated interactions \cite{tserkovRMP05}. In Sec.~\ref{fa}, we specialize to the ferromagnetic case, first revealing EPs in simple two-spin dynamics and then extending the treatment to the magnon band structure in a spin chain. In Sec.~\ref{aa}, a similar program is carried out for the antiferromagnetic case, before summarizing the paper in Sec.~\ref{so}.

\section{General two-spin dynamics}
\label{gtsd}

\subsection{Reactive and dissipative coupling}

Consider an isotropic system composed of two classical spins described by the Hamiltonian
\eq{
H=-\bb_1\cdot\bS_1-\bb_2\cdot\bS_2-J\bS_1\cdot\bS_2\,.
\label{H}}
$\bb_\i$ parametrize individual Zeeman splittings and $J$ Heisenberg exchange between the spins. This Hamiltonian, according to the classical spin algebra $\{S^a,S^b\}=\epsilon^{abc}S_c$ (with $\{\dots\}$ standing for the Poisson bracket and $\epsilon^{abc}$ denoting the Levi-Civita symbol), describes a coupled Larmor precession of the spins:
\eq{
\dot{\bS}_\i\equiv\{\bS_\i,H\}=\bS_\i\times\left(\bb_\i+J\bS_{\tilde{\i}}\right)\,.
}
Here, $\tilde{\i}=2,1$ for $\i=1,2$, respectively. A possible physical realization of such a system can be provided by a magnetic bilayer coupled through a normal-metal spacer \cite{heinrichPRL03}.

\begin{figure}[!h]
\includegraphics[width=0.65\linewidth]{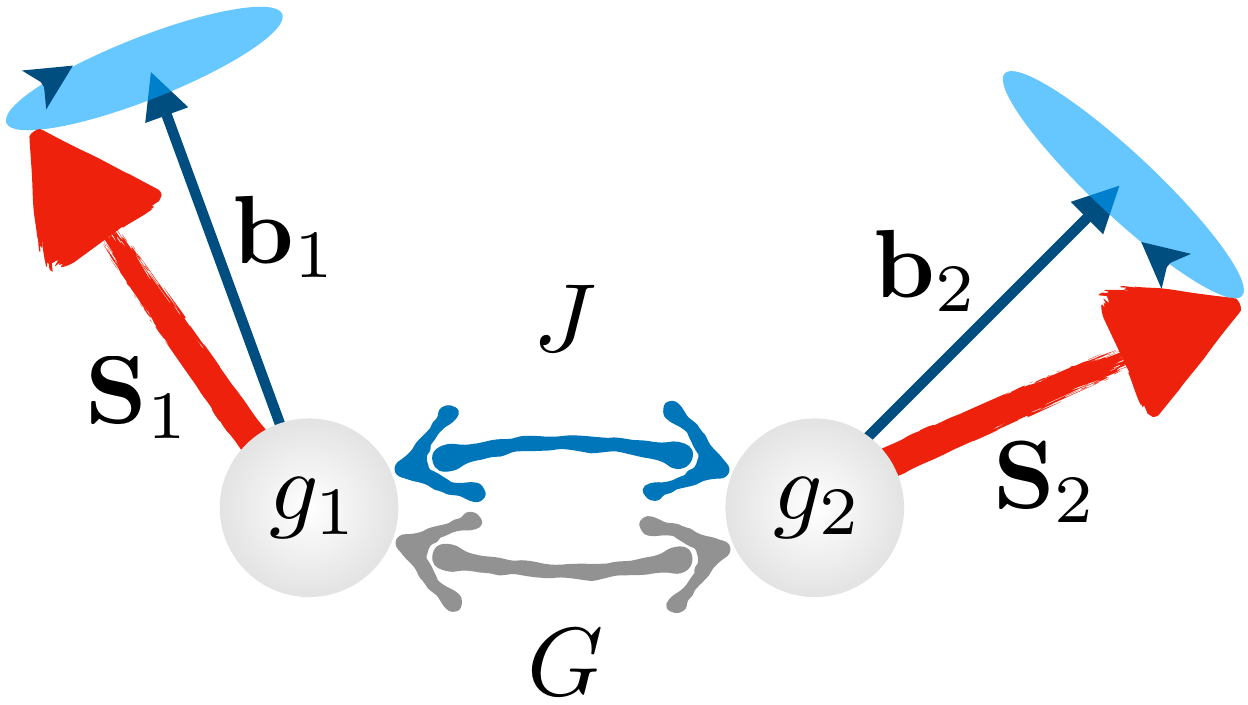}
\caption{A general two-spin system coupled through a dissipative environment. The spins $\bS_\i$ precess in their respective internal fields $\bb_\i$ and couple through a Heisenberg exchange $J$. In addition, individual viscous (Gilbert) damping $g_\i$ is complemented with a mutual spin-pumping-mediated viscosity $G$ in the coupled dynamics.}
\label{two}
\end{figure}

In addition to a RKKY-type exchange $J$, we generally also need to add a dissipative coupling mediated by spin pumping through the spacer \cite{heinrichPRL03,*tserkovPRB03sv}, which enters the equations of motion as a nonlocal Gilbert damping \cite{gilbertIEEEM04}. For small-angle dynamics, relative to a common equilibrium orientation (supposing $\bb_\i$ are collinear), the full coupled equations become:
\eq{
\left(1+g_\i\bS_\i\times\right)\dot{\bS}_\i+G (\bs_\i\times\dot{\bs}_\i-\bs_{\tilde{\i}}\times\dot{\bs}_{\tilde{\i}})=\bS_\i\times\left(\bb_\i+J\bS_{\tilde{\i}}\right)\,.
\label{eom}}
$G$ parametrizes the strength of the spin pumping across the spacer (which is related to the spin-mixing conductance \cite{tserkovPRL02sp,tserkovRMP05}), driven by the orientational dynamics of $\bs_\i\equiv\bS_\i/S_\i$. We also included local Gilbert damping \cite{gilbertIEEEM04} $g_\i$ in each of the magnetic layers, which parametrizes the quality factor $Q_\i=1/2g_\i S_\i$ of intrinsic magnetic dynamics. See Fig.~\ref{two} for a schematic. Note that the equations of motion \rf{eom} would need to be revised in the general large-angle case \cite{tserkovPRB03sv,tserkovRMP05}, in order to preserve the magnitude of $\bS_\i$. In the following, we will see that the nonlocal damping $\propto G$ is essential in establishing EPs in the ferromagnetic case, while it will turn out to be unimportant for the antiferromagnetic case.

\subsection{Energy dissipation and external pumping}

Note that the dissipative coupling $\propto G$ affects only the out-of-phase precession of the two (macro)spin orientations. This could be thought of as a viscosity associated with relative spin dynamics \cite{heinrichPRL03}. Calculating energy dissipation according to the equation of motion \rf{eom},
\eq{
P\equiv-\dot{H}=\bH_1\cdot\dot{\bS}_1+\bH_2\cdot\dot{\bS}_2\,,
}
where $\bH_\i\equiv-\p_{\bS_\i} H$ is the effective field conjugate to spin $\i$, we find
\eq{
P=g_1\dot{\bS}_1^2+g_2\dot{\bS}_2^2+G(\dot{\bs_1}\mp\dot{\bs}_2)^2\,,
\label{P}}
in the case of the dynamics near the parallel (antiparallel) configuration. This dissipation is guaranteed to be positive-semidefinite in the physical situation where $g_1,g_2,G\geq0$ \cite{tserkovPRL02sp}. The above equations thus describe the dynamics of a stable system near its thermodynamic equilibrium.

Formally, $P$ is positive-semidefinite iff $g_\i+G/S_\i^2\geq0$ and $|G|/S_1S_2\leq\sqrt{(g_1+G/S_1^2)(g_2+G/S_2^2)}$. Macroscopic magnetic spins can be locally (thermoelectrically) pumped out of their natural thermodynamic equilibrium by subjecting them to spin-transfer \cite{slonczewskiJMMM96,*bergerPRB96} or spin Seebeck \cite{bauerNATM12} torques, which can shift the effective local damping $g_\i$ into negative values \cite{benderPRL12,*benderPRB14} and possibly even invalidate the stability requirement $P\geq0$. We can exploit this in practice for expanding the parameter space of experimentally tunable coefficients that govern our dynamical system.

\section{Ferromagnetic alignment}
\label{fa}

To be more specific, we now orient the Zeeman terms along the $z$ axis, $\bb_\i=b_\i\bz$, and linearize spin dynamics in terms of small deviations from the initially parallel configuration, $\bS_\i\approx S_\i\bz$. Owing to the axial symmetry, we switch to the natural circular coordinates: $\fS_\i\equiv(S^x_\i+iS^y_\i)/\sqrt{2S_\i}$, which obey canonical algebra (in the case of small-angle dynamics):
\eq{
i\{\fS,\fS^*\}\approx1\,,
}
for each site $\i$ (with the inter-site Poisson brackets vanishing). We thus see that the quantized $\fS\to\sqrt{\hbar}a$ obey bosonic statistics: $[a,a^\dagger]\to i\{\fS,\fS^*\}=1$. $a$, which is proportional to the spin-raising operator, thus constitutes the magnon field.

\subsection{Equation of motion}

The linearized dynamics following from Eq.~\rf{eom} are described by
\eq{
(1+i\alpha_\i)\dot{\fS}_\i-i\alpha'\dot{\fS}_{\tilde{\i}}=-i\omega_\i\fS_\i+i\omega'\fS_{\tilde{\i}}\,,
\label{dsi}}
where $\omega_\i\equiv b_\i+JS_{\tilde{\i}}$, $\alpha_\i\equiv g_\i S_\i+G/S_\i$ and we denoted by the primed coefficients, $\omega'\equiv J\sqrt{S_1S_2}$ and $\alpha'\equiv G/\sqrt{S_1S_2}$, the reactive and dissipative inter-spin couplings, respectively. Equation \rf{dsi} can finally be recast in the matrix (Schr{\"o}dinger-like) form:
\eq{
(1+i\hat{d})i\frac{d}{dt}\fS=\hat{h}\fS\,,
\label{dGh}}
where
\eq{
\fS=\left(\ar{c}{\fS_1\\\fS_2}\right)\,,~~~\hat{h}=\omega_++\omega_-\hat{\sigma}_z-\omega'\hat{\sigma}_x
\label{fS}}
is the effective $2\times2$ magnon Hamiltonian, and
\eq{
\hat{d}=\alpha_++\alpha_-\hat{\sigma}_z-\alpha'\hat{\sigma}_x
}
is the damping tensor, parametrized by the Pauli matrices $\hat{\sigma}_a$. $\omega_\pm\equiv(\omega_1\pm \omega_2)/2$ (and similarly for $\alpha_\pm$) are the (anti)symmetrized frequencies. The dynamics is finally cast as
\eq{
i\frac{d}{dt}\fS=\hat{H}\fS\,,
}
in terms of the non-Hermitian ``Hamiltonian"
\eq{
\hat{H}\equiv(1+i\hat{d})^{-1}\hat{h}\,.
\label{nH}}

Let us simplify $\hat{H}$ by assuming smallness of the damping parameters, $\alpha\ll1$ (implying large quality factors for the resonant modes), as well as of the interspin  detuning and coupling, $\omega_-,\omega'\ll\omega_+$. This leads to
\eq{\al{
\hat{H}&\approx(1-i\alpha_-\hat{\sigma}_z+i\alpha'\hat{\sigma}_x)(1+\gamma_-\hat{\sigma}_z-\gamma'\hat{\sigma}_x)\tilde{\omega}_+\\
&\approx\left[1+(\gamma_--i\alpha_-)\hat{\sigma}_z-(\gamma'-i\alpha')\hat{\sigma}_x\right]\tilde{\omega}_+\\
&\equiv(1+\FH)\tilde{\omega}_+\approx\tilde{\omega}_++\FH\omega_+\,,
\label{Hpl}}}
where $\gamma_-\equiv\omega_-/\omega_+\ll1$ and $\gamma'\equiv\omega'/\omega_+\ll1$ are the normalized detuning and coupling, respectively, and $\tilde{\omega}_+\equiv\omega_+/(1+i\alpha_+)$ is the complex-valued normal frequency of the unperturbed symmetric mode. By diagonalizing the normalized and shifted Hamiltonian
\eq{
\FH=(\gamma_--i\alpha_-)\hat{\sigma}_z-(\gamma'-i\alpha')\hat{\sigma}_x\,,
\label{fH}}
we can finally decompose the linearized dynamics into damped modes of the form $\propto e^{-i(1+\lambda)\tilde{\omega}_+t}$, where $\lambda$ is one of the two (complex-valued) eigenvalues of $\FH$. For convenience, we summarize the key quantities parametrizing the coupled dynamics in Table~\ref{tab}.
\begin{table}[h]
\caption{Parameters of the collinear two-spin system.}
\begin{tabular}{cl}
\hline\hline
 $\gamma_-$ & normalized frequency asymmetry, $\omega_-/\omega_+$\\
 $\alpha_-$ & damping asymmetry\\
 \hline
 $\gamma'$ & normalized exchange coupling, $\omega'/\omega_+$\\
 $\alpha'$ & dissipative coupling\\
\hline
\end{tabular}
\label{tab}
\end{table}

\subsection{Exceptional points}

The diagonalization of the matrix \rf{fH} breaks down at the exceptional points, where the eigenvectors associated with degenerate eigenvalues coalesce \cite{heissPRE00}, ruling out a diagonalizing similarity transformation. The two eigenvalues are given by
\eq{
\lambda_\pm=\pm\sqrt{(\gamma_--i\alpha_-)^2+(\gamma'-i\alpha')^2}\,,
\label{l}}
where we are making the convention for the square root to evaluate the principal value. An EP occurs when $\lambda_\pm=0$, while the individual terms under the square root are not both zero. In this case, $\FH\neq0$, while $\FH^2=0$, which confirms that indeed the matrix cannot be diagonalized. A ready example of this is provided by a matrix $\propto\hat{\sigma}_a+i\hat{\sigma}_b$, in terms of two distinct Pauli matrices $\hat{\sigma}_a$ and $\hat{\sigma}_b$, as we have already mentioned.

Expanding the (complex-valued) energy eigenvalues \rf{l} near such an EP, as a function of some complex-valued parameter that parametrizes $\FH$, would generically define a square-root singularity. The trivial degeneracy of $\FH=0$ signals a \textit{diabolic point}, on the other hand, which would harbor a Berry-curvature monopole \cite{berryPRSLA84} associated with a Weyl singularity.

\subsubsection{The reactive and dissipative scenarios}

Even if the two-spin system is symmetric, $S_1=S_2$, we may still introduce possible asymmetries in the individual resonant frequencies and the associated broadenings. The EP condition \rf{l} translates into
\eq{
\gamma_--i\alpha_-=\pm i(\gamma'-i\alpha')\neq0\,.
}
The EPs can then be realized for physical real-valued parameters when $\alpha_-=\mp\gamma'$ and $\gamma_-=\pm\alpha'$. Two basic practical scenarios can then be envisioned: (1) The coupling between the spins is purely reactive, $\gamma'\neq0$, while $\alpha'=0$, in which case the resonances need to be tuned, $\gamma_-=0$, resulting in two EPs when $\alpha_-=\pm\gamma'$; and (2) The coupling between the spins is purely dissipative, $\alpha'\neq0$, while $\gamma'=0$, in which case the local dissipation needs to be symmetric, $\alpha_-=0$, resulting in two EPs when $\gamma_-=\pm\alpha'$. The latter scenario is especially attractive, as it naturally occurs in a magnetic bilayer system with a diffusive normal-metal spacer \cite{tserkovPRB03sv,tserkovRMP05}. Resonant tuning across the EP has been realized in Ref.~\cite{heinrichPRL03}, not making the connection with the EP perspective, however, at the time.

\subsubsection{Topology of the exceptional points}

\begin{figure}[!h]
\includegraphics[width=0.75\linewidth]{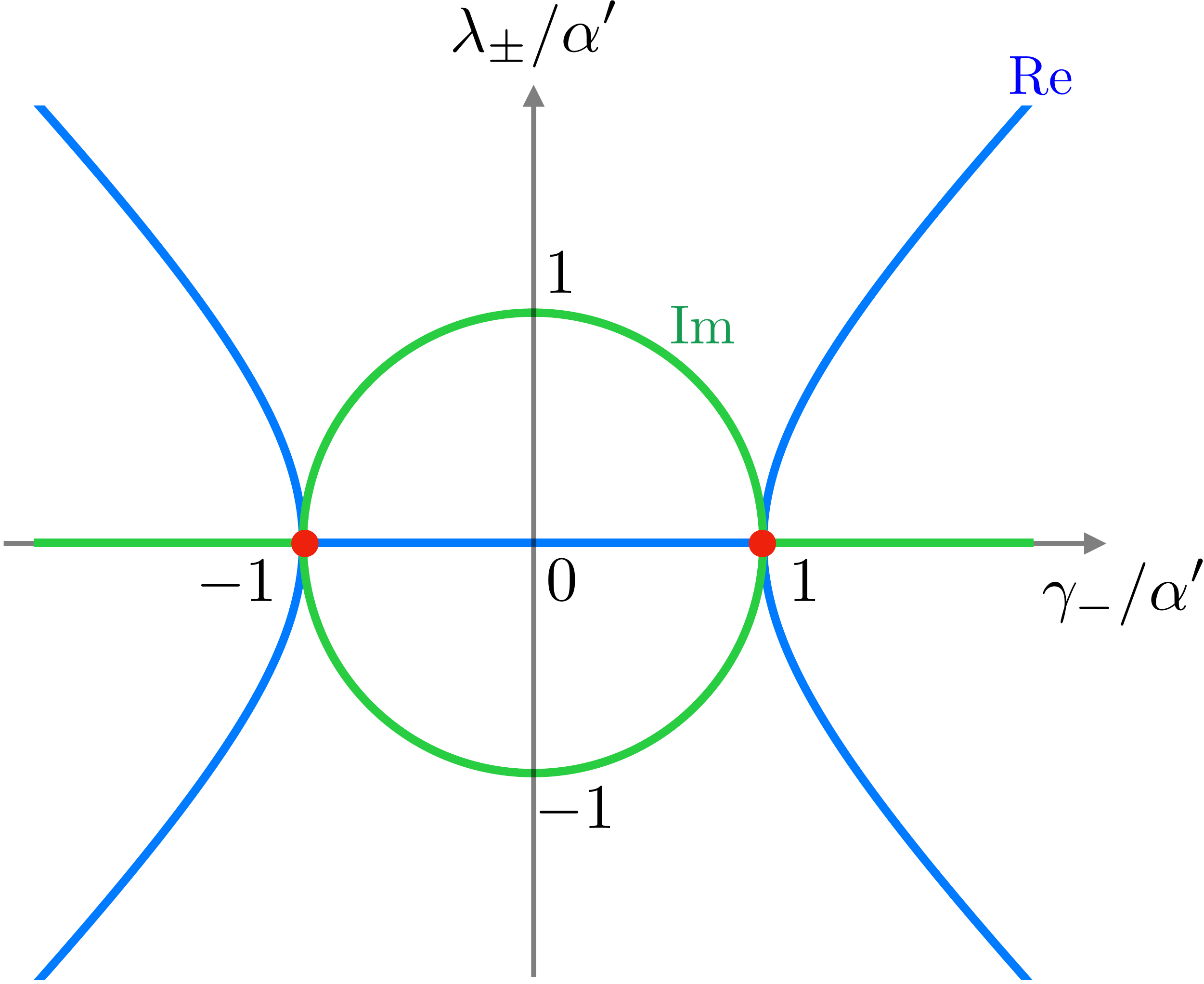}
\caption{Complex-valued eigenfrequencies \rf{ld}, when $\alpha_-=0$. The two EPs at $|\gamma_-/\alpha'|=1$ engender singularities, at which the eigenfrequencies switch from being purely imaginary (at smaller local frequency asymmetries $\gamma_-$) to purely real (at larger asymmetries $\gamma_-$). Such eigenfrequency structure has been observed in a symmetric iron-based magnetic bilayer \cite{heinrichPRL03} (see also Ref.~\cite{tserkovRMP05}, for a more detailed analysis), as well as hybrid microwave-cavity based systems \cite{zhangNATC17,bernierPRA18,harderPRL18}.}
\label{sym}
\end{figure}

Let us look more closely into the frequency eigenvalues \rf{l} in the vicinity of these exceptional points. As an example, suppose we have a magnetic bilayer coupled through a purely dissipative coupling, i.e., $\gamma'=0$. Let us take the dissipative coupling $\alpha'$ to be fixed, while the local resonance conditions and damping, $\gamma_-$ and $\alpha_-$, are allowed to be tuned by the control of local fields and thermoelectric pumping \cite{benderPRB14}. The frequency eigenvalues are then given by
\eq{
\lambda_\pm=\pm\sqrt{(\gamma_--i\alpha_-)^2-\alpha'^2}\,.
\label{ld}}
We can rewrite these eigenfrequencies as $\lambda_\pm=\pm\sqrt{z^2-\alpha'^2}$, in terms of the fully-tunable complex-valued parameter $z\equiv\gamma_--i\alpha_-$ [which couples to the $z$-component Pauli matrix in the original Hamiltonian \rf{fH}]. If $\alpha_-=0$, $z$ is swept along the real axis by varying $\gamma_-$, passing through two EP at $z=\pm\alpha'$. The corresponding frequency eigenvalues are plotted in Fig.~\ref{sym}. For $|\gamma_-/\alpha'|<1$, they are purely imaginary, with the antisymmetric mode damped relative to the symmetric one. Amusingly, both modes here are perfectly synchronized, in regard to their real frequency components. This is dubbed level attraction \cite{bernierPRA18,harderPRL18}, in contrast to the usual level repulsion for a hybridized Hermitian system. At $|\gamma_-/\alpha'|>1$, the modes bifurcate, with a vanishing relative damping. The EP transition from the purely real to the purely imaginary eigenvalues $\lambda_\pm$ at $|\gamma_-/\alpha'|=1$, as depicted in Fig.~\ref{sym}, is related to the spontaneous breaking of a $\mathcal{PT}$ symmetry \cite{benderRPP07}.

It may be useful to recall, that these frequency eigenvalues $\lambda_\pm$ are normalized and shifted by a complex-valued $\tilde{\omega}_+$ [cf. Eq.~\rf{Hpl}]. The corresponding physical eigenfrequencies must both have negative imaginary components, according to the overall thermodynamic stability of the system [so long as dissipation $P$ in Eq.~\rf{P} is positive, which should be true in and near equilibrium].

Close to either of the two EPs, $z\to z_0=\pm\alpha'$, we obtain a square-root singularity in the full complex plane, $z\in\mathbb{C}$:
\eq{
\lambda_\pm\to\pm\sqrt{2z_0(z-z_0)}~~~(\textrm{when}~\alpha'\neq0)\,.
\label{sr}}
When the dissipative coupling vanishes, $\alpha'\to0$, on the other hand, the two EPs merge, resulting in a single Weyl point in the spectrum (at $\FH\to0$):
\eq{
\lambda_\pm\to\pm z~~~(\textrm{when}~\alpha'=0)\,.
}
The latter is of course just a trivial scenario of decoupled circular precession, near the degeneracy point. For a finite spin pumping, $\alpha'\neq0$, the eigenfrequencies $\propto\sqrt{z}$ constitute a double-valued function in the original complex plane $z$, while being single-valued on the Riemann surface, which consists of two sheets emanating from the EP (the branch point) and stitched up along the branch cut (that is typically chosen along the negative real axis of $z$).

\begin{figure}[!h]
\includegraphics[width=0.8\linewidth]{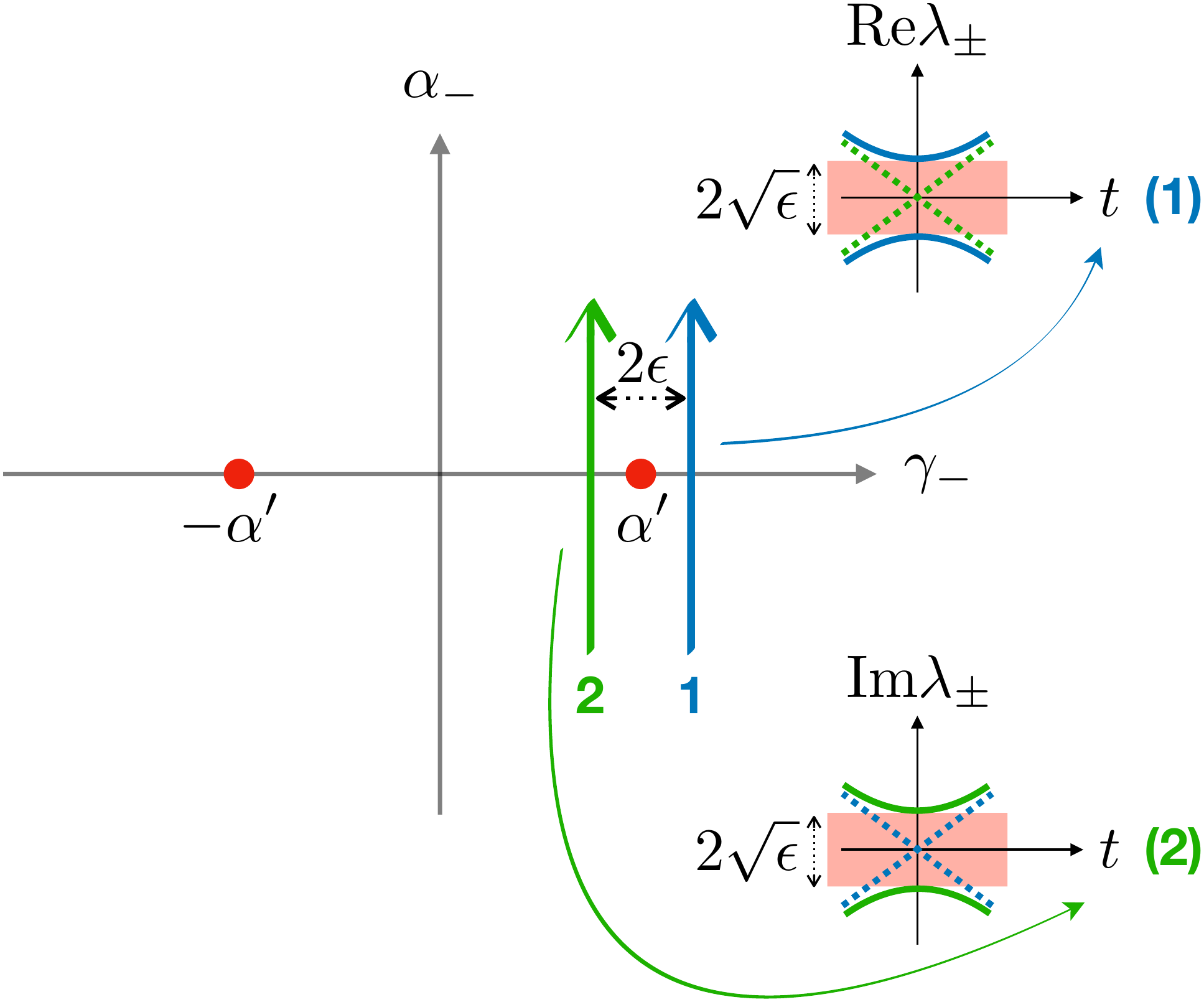}
\caption{The evolution of the eigenvalues upon the passage of an exceptional point $z_0\to\alpha'$ to the right (1) or left (2). The solid lines in the insets show the respective values of the real and imaginary parts of the frequency eigenvalues (in units of $\sqrt{2\alpha'}$). The dashed lines are, respectively, the imaginary and real parts.}
\label{paths}
\end{figure}

If $z$ passes just to the right of the EP (path $1$ in Fig.~\ref{paths}): $z=z_0+\epsilon-it$, where $\epsilon>0$ is a small shift controlled by fixing $\gamma_-\to z_0+\epsilon$ and $\alpha_-\to t$ is continuously varied, the eigenvalues
\eq{
\lambda_\pm\propto\pm\sqrt{\epsilon-it}
}
have their two real parts anticrossing and two imaginary parts crossing, at $t\to0$. When the passage is performed to the left of the EP (path $2$ in Fig.~\ref{paths}), $\epsilon<0$, the real parts cross while the imaginary parts anticross. At the Weyl point, $\epsilon=0$, both parts cross, of course. This crossing/anticrossing behavior is generic for basic topological reasons \cite{heissPRE00}.

\subsection{Spin chain}

The EPs discussed above can also be approached in momentum space, by considering a dimerized chain of two-spin composites. As a starting point to that end, we analyze a chain of $N$ identical spins, with periodic boundary conditions, exchange-coupled as in Eq.~\rf{H}:
\eq{
H=-\sum_{\j=1}^N\bb_\j\cdot\bS_\j-J\sum_{\langle \j\j'\rangle}\bS_\j\cdot\bS_{\j'}\,,
}
where the double sum runs over the nearest neighbors (the $N$th spin being a neighbor to the 1st one). The linearized dynamics \rf{dsi} then obey
\eq{\al{
(1+ig_\j S)\dot{\fS}_\j-&i\alpha'(\dot{\fS}_{\j-1}+\dot{\fS}_{\j+1}-2\dot{\fS}_\j)/2\\
&=-ib_\j\fS_\j+i\omega'(\fS_{\j-1}+\fS_{\j+1}-2\fS_\j)/2\,,
}}
scaling, for convenience, the definitions of the coupling parameters $\alpha'$ and $\omega'$ up by a factor of 2. If $g_\j\equiv\alpha/S$ and $b_\j\equiv b$, the solutions are plane waves $\fS_j\propto e^{i(kj-\omega t)}$, with the dispersion
\eq{
\omega=\frac{b+2\omega'\sin^2\frac{k}{2}}{1+i\left(\alpha+2\alpha'\sin^2\frac{k}{2}\right)}\,.
\label{wsc}}
It has a finite positive curvature in the (real part of the) frequency as well as in the effective Gilbert damping (or the inverse quality factor), as $k\to0$ \cite{tserkovRMP05}. The wave number $k$ runs over the Brillouin zone $(-\pi,\pi]$. Assuming the linearization of the spin dynamics is performed with respect to a stable state (i.e., energy minimum of an isolated spin chain), the eigenfrequency \rf{wsc} exhibits no singularities.

\begin{figure}[!h]
\includegraphics[width=0.8\linewidth]{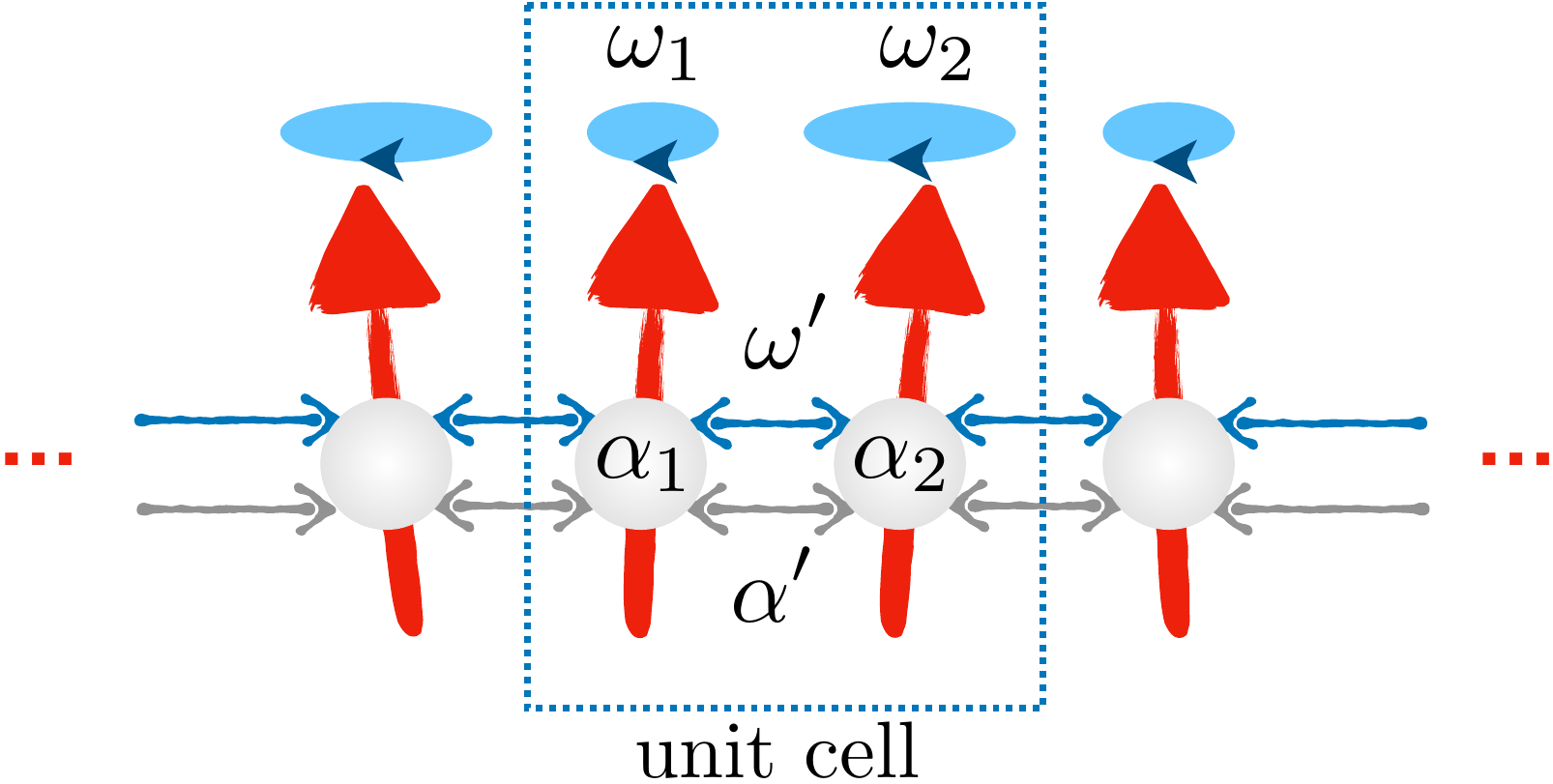}
\caption{Spin waves in a ferromagnetically-ordered spin chain. The neighboring sites are interacting via reactive, $\omega'$, and dissipative, $\alpha'$, exchange couplings. A unit cell is composed of a nonidentical spin pair, with the individual frequencies and damping parametrized by $\omega_\i$ and $\alpha_\i$, respectively.}
\label{f}
\end{figure}

To allow for a dimerization of spin dynamics (a term we use broadly to account for a two-spin unit cell), we next suppose the onsite field and damping are alternating in magnitude between neighboring sites, as in the preceding two-spin model. See Fig.~\ref{f} for a schematic. We will now look for solutions of the form
\eq{
\fS_{\i\j}(t)=\fS_\i(t)e^{ik\j}\,,
\label{Gij}}
where $\i=1,2$ is the sublattice index and $\j$ is the unit-cell index labeling repeated site pairs. The ensuing dynamics of $\fS_\i(t)$ are then governed by a $k$-dependent $2\times2$ Hamiltonian:
\eq{\al{
\hat{H}=&\left[1+i\left(\alpha_++\alpha_-\hat{\sigma}_z-\alpha'\cos\frac{k}{2}\hat{\sigma}_x^k\right)\right]^{-1}\nonumber\\
&\times\left(\omega_++\omega_-\hat{\sigma}_z-\omega'\cos\frac{k}{2}\hat{\sigma}_x^k\right)\,.
}}
This is fully analogous to Eqs.~\rf{fS}-\rf{nH} but with $\alpha'\to\alpha'\cos\frac{k}{2}$ and $\omega'\to\omega'\cos\frac{k}{2}$ now modulated by the factor of $\cos\frac{k}{2}$ and rotating the Pauli matrix,
\eq{
\hat{\sigma}_x\to\hat{\sigma}_x^k\equiv\cos\frac{k}{2}\hat{\sigma}_x+\sin\frac{k}{2}\hat{\sigma}_y\,,
}
by the angle $k/2$ around the $z$ axis.

Focusing on a purely dissipative coupling by setting $\omega'\to0$, as before, we find frequency eigenvalues [normalized by $\tilde{\omega}_+\equiv\omega_+/(1+i\alpha_+)$] in the form \rf{ld}:
\eq{
\lambda_\pm=\pm\sqrt{(\gamma_--i\alpha_-)^2-\alpha'^2\cos^2\frac{k}{2}}\,,
\label{lSC}}
with $\gamma_-$ and $\alpha_-$ now denoting the frequency and damping asymmetries on the adjacent lattice sites. The role of the spin-pumping coupling $\alpha'$ is maximized at $k=0$, where the antisymmetric mode gets damped relative to the symmetric one, and gets diminished towards the Brillouin zone boundaries.

\begin{figure}[!h]
\includegraphics[width=\linewidth]{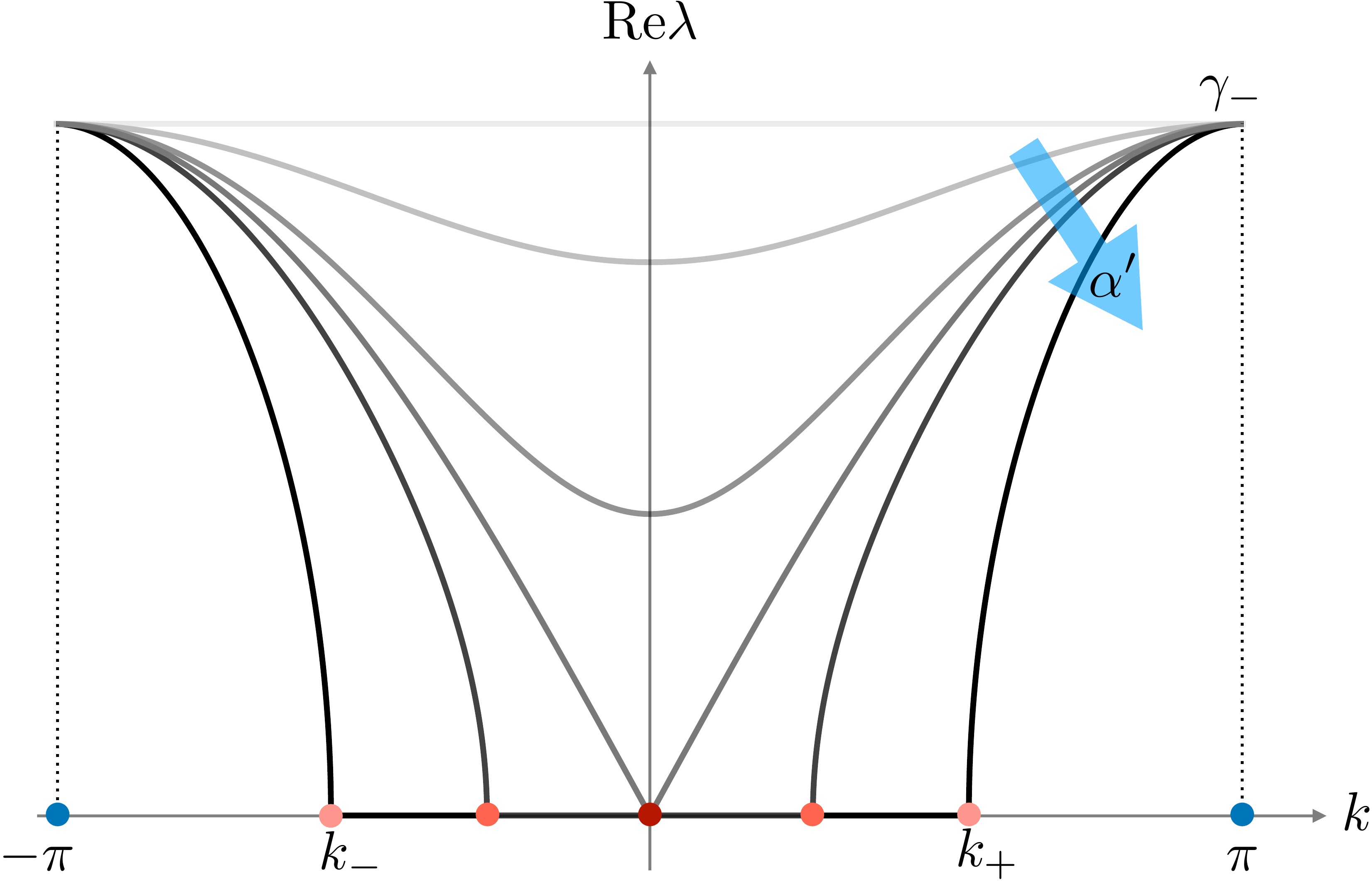}
\caption{The (positive real part of the) eigenfrequencies \rf{lSC}, with $k$ swept over the Brillouin zone. Here, we set $\alpha_-\to0$ and vary $\alpha'/\gamma_-\to\{0,0.6,0.9,1,1.1,1.5\}$. The EPs $k_\pm$, Eqs.~\rf{kpm}, are marked for $\alpha'/\gamma_-=1.5$. ${\rm Im}\,\lambda=0$ when ${\rm Re}\,\lambda\neq0$ and vice versa.}
\label{bz}
\end{figure}

As before, we can reach the EP singularity by setting $\alpha_-\to0$ and varying $\gamma_-$ and/or $k$. Physically, this corresponds to a homogeneous spin chain, apart from a staggering of the applied magnetic field (parametrized by $\gamma_-$). Suppose all the parameters of the system are fixed, with $|\alpha'/\gamma_-|>1$, so that there are two real-valued momenta,
\eq{
k_\pm=\pm2\cos^{-1}\left|\frac{\gamma_-}{\alpha'}\right|\,,
\label{kpm}}
at which the EPs shown in Fig.~\ref{sym} are traversed as we move within the Brillouin zone. For $k\in(k_-,k_+)$, we are effectively inside the circle shown in Fig.~\ref{sym}, while outside otherwise. In Fig.~\ref{bz}, we plot the corresponding (real part of the) eigenfrequencies, $\lambda_\pm/\gamma_-$, for $\alpha'/\gamma_-$ varied between $0$ and $5$. $\alpha'/\gamma_-\to1$ corresponds to the critical point, at which the two EPs merge into a single Weyl point and subsequently disappear from the real momentum axis at smaller dissipative coupling $\alpha'$.

Note that at $\alpha'/\gamma_-<1$, a finite dissipative coupling $\alpha'$ endows dynamics with a dispersion, despite the absence of any Heisenberg exchange $J$. For $\alpha'/\gamma_->1$, on the other hand, the two modes get synchronized at ${\rm Re}\,\lambda_\pm=0$ (with one mode damped relative to the other) for $k_-<k<k_+$. These two qualitatively distinct regimes of the coupled dynamics are separated by a Weyl point emerging at $\alpha'/\gamma_-=1$. We expect these characteristic dispersions, synchronization, and Weyl criticality to provide practical experimental handles to explore the consequences of the emergence of the EPs in our spin chain. It is intriguing, in particular, how the group velocity changes in a steplike fashion around the EPs $k_\pm$.

We remind, in the closing of this section, that these dynamics are associated with an overall decaying envelope function governed by the complex-valued frequency $\tilde{\omega}_+\equiv\omega_+/(1+i\alpha_+)$, which would ensure stability in equilibrium. If the spin chain is thermoelectrically pumped \cite{benderPRL12,*benderPRB14}, in order to effectively tune $\alpha_+$ to zero, the symmetric mode would go unstable within the $(k_-,k_+)$ interval delineated by the EPs, if $\alpha'>|\gamma_-|$.

\section{Antiferromagnetic alignment}
\label{aa}

\subsection{Two spins}

Returning to our two-spin dynamics, Eq.~\rf{eom}, let us now consider an antiferromagnetic (AF) state, $J<0$, of two equal spins. We will suppose $\bS_\i\approx(-1)^\i S\bz$, for sites $\i=1,2$. The local fields are $\bb_\i=[b+(-1)^\i K]\bz=(-1)^\i b_\i\bz$, $b_\i\equiv(-1)^\i b+K$, in terms of an easy-axis anisotropy $K\geq0$ and a collinearly applied uniform field $b$. The canonical transverse coordinates are now conveniently defined as $\fS_\i\equiv(-1)^\i(S^x_\i+iS^y_\i)/\sqrt{2S}$, obeying $i\{\fS_\i,\fS_\i^*\}\approx(-1)^\i$. Linearizing equations of motion \rf{eom} in terms of these coordinates, we get
\eq{
\left[(-1)^\i+i\alpha_\i\right]\dot{\fS}_\i-i\alpha'\dot{\fS}_{\tilde{\i}}=-i\omega_\i\fS_\i+i\omega'\fS_{\tilde{\i}}\,,
\label{mpi}}
where, as before, $\omega_\i\equiv b_\i+\omega'$, $\alpha_\i\equiv g_\i S+\alpha'$, and we denoted by the primed coefficients $\omega'\equiv|J|S$ and $\alpha'\equiv G/S$. Adhering to the form of Eq.~\rf{dGh}, this system of two equations can be written in terms of
\eq{
\hat{d}=-g_-S-(g_+S+\alpha')\hat{\sigma}_z+i\alpha'\hat{\sigma}_y
\label{dAF}}
and
\eq{
\hat{h}=b-(K+\omega')\hat{\sigma}_z+i\omega'\hat{\sigma}_y\,,
\label{hAF}}
where $g_\pm\equiv(g_1\pm g_2)/2$. The net effective non-Hermitian Hamiltonian \rf{nH} is thus
\eq{\al{ 
\hat{H}\approx&\left[1+ig_-S+i(g_+S+\alpha')\hat{\sigma}_z+\alpha'\hat{\sigma}_y\right]\\
&\times\left[b-(K+\omega')\hat{\sigma}_z+i\omega'\hat{\sigma}_y\right]\,,
}}
supposing, as before, that all the dimensionless damping parameters are small.

Setting $g_-\to0$, $g_+S+\alpha'\to\alpha$, $b\to0$, and $K+\omega'\to\kappa$, we get
\eq{\al{
\hat{H}&\approx\left(1+i\alpha\hat{\sigma}_z+\alpha'\hat{\sigma}_y\right)\left(i\omega'\hat{\sigma}_y-\kappa\hat{\sigma}_z\right)\\
&=-i(\alpha\kappa-\alpha'\omega')+i(\alpha\omega'-\alpha'\kappa)\hat{\sigma}_x+(i\omega'\hat{\sigma}_y-\kappa\hat{\sigma}_z)
\,,
\label{HAF}}}
with constraints $\alpha\geq\alpha'$ and $\kappa\geq\omega'$. This Hamiltonian describes a two-site antiferromagnet, with effective local damping $\alpha$, local Larmor frequency $\kappa$, dissipative coupling $\alpha'$, and exchange coupling $\omega'$. Dropping the constant piece, $\propto(\alpha\kappa-\alpha'\omega')\geq0$, which governs an overall decay, we get
\eq{
H^2=\kappa^2-\omega'^2-(\alpha\omega'-\alpha'\kappa)^2\,.
\label{H2AF}}
Setting this to zero, in order to locate the EP, we thus require
\eq{
\alpha\omega'-\alpha'\kappa=\pm\sqrt{\kappa^2-\omega'^2}\,,
}
with the expression under the square root being positive (for a stable AF configuration with $K\geq0$). Writing $\alpha=\fa+\alpha'$, where $\fa\geq0$ is  the intrinsic local damping (excluding spin pumping), the EP condition can be rewritten as
\eq{
\fa\omega'=\alpha' K\pm\sqrt{K(K+2\omega')\,.}
}
Supposing, as before, that $\alpha'\ll1$ and also $\omega'\gg K$ (strong exchange), we finally get
\eq{
\fa\approx\sqrt{\frac{2K}{\omega'}}\ll1\,,
}
which is consistent with the smallness of $\fa$. This value of damping corresponds to the quality factor of the AF resonance $\sim1$, however. In the spirit of these approximations, we thus end up with the full frequency eigenvalues following from Hamiltonian \rf{HAF}:
\eq{
\lambda_\pm\approx-i\fa\omega'\pm\sqrt{2(1-\fa\omega'/\omega)}\omega\,,
}
near the EP point $\fa\approx\omega/\omega'$, where $\omega\equiv\sqrt{2K\omega'}$ is the intrinsic AF resonance frequency.

\subsection{AF vs F cases}
\label{AFF}

It is amusing to remark that two spins interacting by a pure antiferromagnetic exchange, in the absence of any damping and additional fields, naturally realize an EP:
\eq{
\hat{H}_{\rm AF}'=\omega'(i\hat{\sigma}_y-\hat{\sigma}_z)\,,
}
according to Eqs.~\rf{dAF} and \rf{hAF}, after setting to zero all terms but $\omega'$. This is in stark contrast to the analogous ferromagnetic case, where
\eq{
\hat{H}_{\rm F}'=\omega'(1-\hat{\sigma}_x)\,,
}
according to Eq.~\rf{fS}. $\hat{H}_{\rm AF}'^2=0$, while  $(\hat{H}_{\rm F}'/\omega'-1)^2=1$ (having subtracted the constant part), suggests that the AF dynamics is more peculiar. Indeed, decomposing the small-angle dynamics into the symmetric and antisymmetric components: $\fS_\pm=(\fS_1\pm\fS_2)/2$, we get
\eq{\al{
{\rm AF:}~~~\dot{\fS}_-&=0\,,\,\,\,\dot{\fS}_+=2i\omega'\fS_-\,,\\
{\rm F:}~~~\dot{\fS}_+&=0\,,\,\,\,\dot{\fS}_-=-2i\omega'\fS_-\,.
}}
Both cases exhibit a zero mode, $\fS_+$, corresponding to a reorientation of the overall order parameter (N{\'e}el in the AF and magnetic in the F cases). The distortion of this order, i.e., $\fS_-\neq0$, triggers its small-angle precession with frequency $2\omega'$ in the F case, while resulting in an apparently unbounded growth of $\fS_+$ in the AF case. The antiferromagnetic EP point thus results in a breakdown of the linearized treatment. We of course know the corresponding outcome in the full spin dynamics: The N{\'e}el order parameter precesses in the plane perpendicular to the distortion $\fS_-$, which parametrizes relative spin canting, with the frequency $\propto\omega'\fS_-$. This simple example illustrates how an EP takes the coupled dynamics out of the linearized perturbative treatment, necessitating a fall back on a more complete description.

\subsection{Spin chain}

\begin{figure}[!h]
\includegraphics[width=0.8\linewidth]{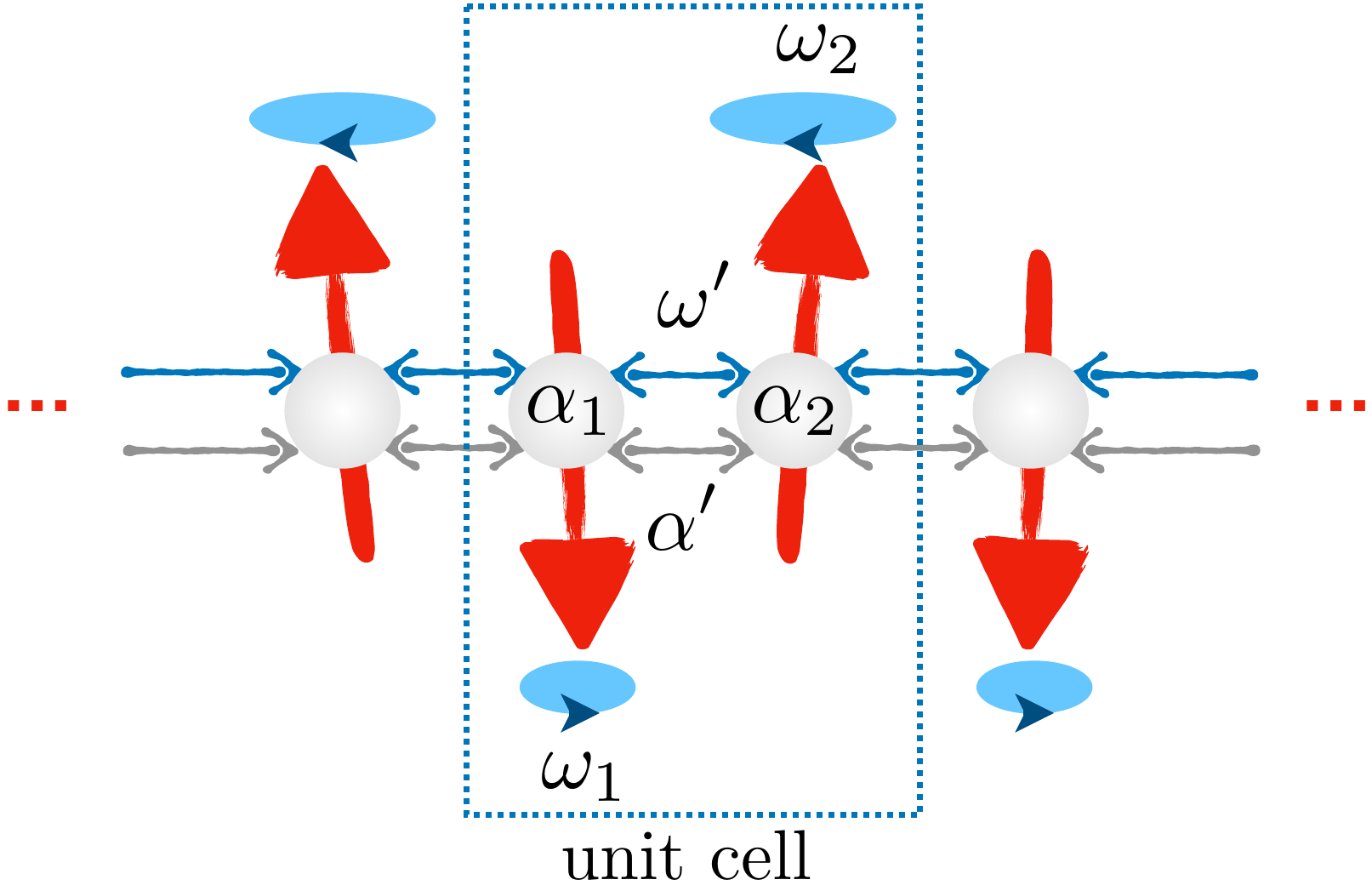}
\caption{Spin waves in an antiferromagnetically-ordered spin chain. The neighboring sites are interacting via reactive, $\omega'$, and dissipative, $\alpha'$, exchange coupling. A unit cell is composed of an antiferromagnetic dimer, with the individual frequencies and damping parametrized by $\omega_\i$ and $\alpha_\i$, respectively.}
\label{af}
\end{figure}

Viewing the above two-spin system as a unit cell of an infinite homogeneous spin chain (which is thus naturally dimerized), we are looking for solutions of the form \rf{Gij}, where $\i=1,2$ is the sublattice index, as before, and $\j$ labels the unit cells. See Fig.~\ref{af} for a schematic. The resultant equations of motion [cf. Eq.~\rf{mpi}] are
\eq{\al{
\left[(-1)^\i+i\alpha_\i\right]\dot{\fS}_\i-&i\alpha'\dot{\fS}_{\tilde{\i}}\frac{1+e^{(-1)^\i ik}}{2}=\\
&-i\omega_\i\fS_\i+i\omega'\fS_{\tilde{\i}}\frac{1+e^{(-1)^\i ik}}{2}\,,
}}
scaling, for convenience, the definitions of the coupling parameters $\alpha'$ and $\omega'$ up by a factor of 2. The analogs of Eqs.~\rf{dAF} and \rf{hAF} become (having set $g_-\to0$, $g_+S+\alpha'\to\alpha$, $b\to0$, and $K+\omega'\to\kappa$):
\eq{
\hat{d}=-\alpha\hat{\sigma}_z+i\alpha'\cos\frac{k}{2}\hat{\sigma}_y^k
}
and
\eq{
\hat{h}=-\kappa\hat{\sigma}_z+i\omega'\cos\frac{k}{2}\hat{\sigma}_y^k\,,
}
where
\eq{
\hat{\sigma}_y^k\equiv\cos\frac{k}{2}\hat{\sigma}_y-\sin\frac{k}{2}\hat{\sigma}_x
}
is the Pauli matrix $\hat{\sigma}_y$ rotated by angle $k/2$ around the $z$ axis. These equations reduce to Eqs.~\rf{dAF} and \rf{hAF} in the limit of $k=0$.

The spectrum and the subsequent EP analysis can thus be obtained from Eq.~\rf{H2AF}, after scaling $\alpha'$ and $\omega'$ by $\cos\frac{k}{2}$, which results in the $k$-dependent eigenfrequencies
\eq{
\lambda_\pm\approx-i\fa\omega'\pm\sqrt{\omega^2+\omega'^2\left(\sin^2\frac{k}{2}-\fa\cos^2\frac{k}{2}\right)}\,.
\label{lAF}}
We are omitting here terms $\propto\alpha'$, which can be shown to be unimportant when the intrinsic damping $\fa$ is approaching the EP point. The EP is located at
\eq{
\fa\cos\frac{k}{2}\approx\sqrt{(\omega/\omega')^2+\sin^2\frac{k}{2}}\,.
}
For $\fa\ll1$, we thus need to focus on $k\to0$, which gives
\eq{
\fa\approx\sqrt{(\omega/\omega')^2+(k/2)^2}\approx\omega_k/\omega'\,,
}
supposing, as before, that $K\ll\omega'$ and thus $\omega\ll\omega'$. $\omega_k\equiv\sqrt{\omega^2+\omega'^2\sin^2\frac{k}{2}}$ is the intrinsic undamped dispersion. Note that in the absence of damping, $\fa\to0$, the EP is reached at $k\to0$, requiring the absence of any anisotropy, $\omega\to0$. This reproduces the elementary antiferromagnetic EP discussed in Sec.~\ref{AFF}.

\begin{figure}[!h]
\includegraphics[width=\linewidth]{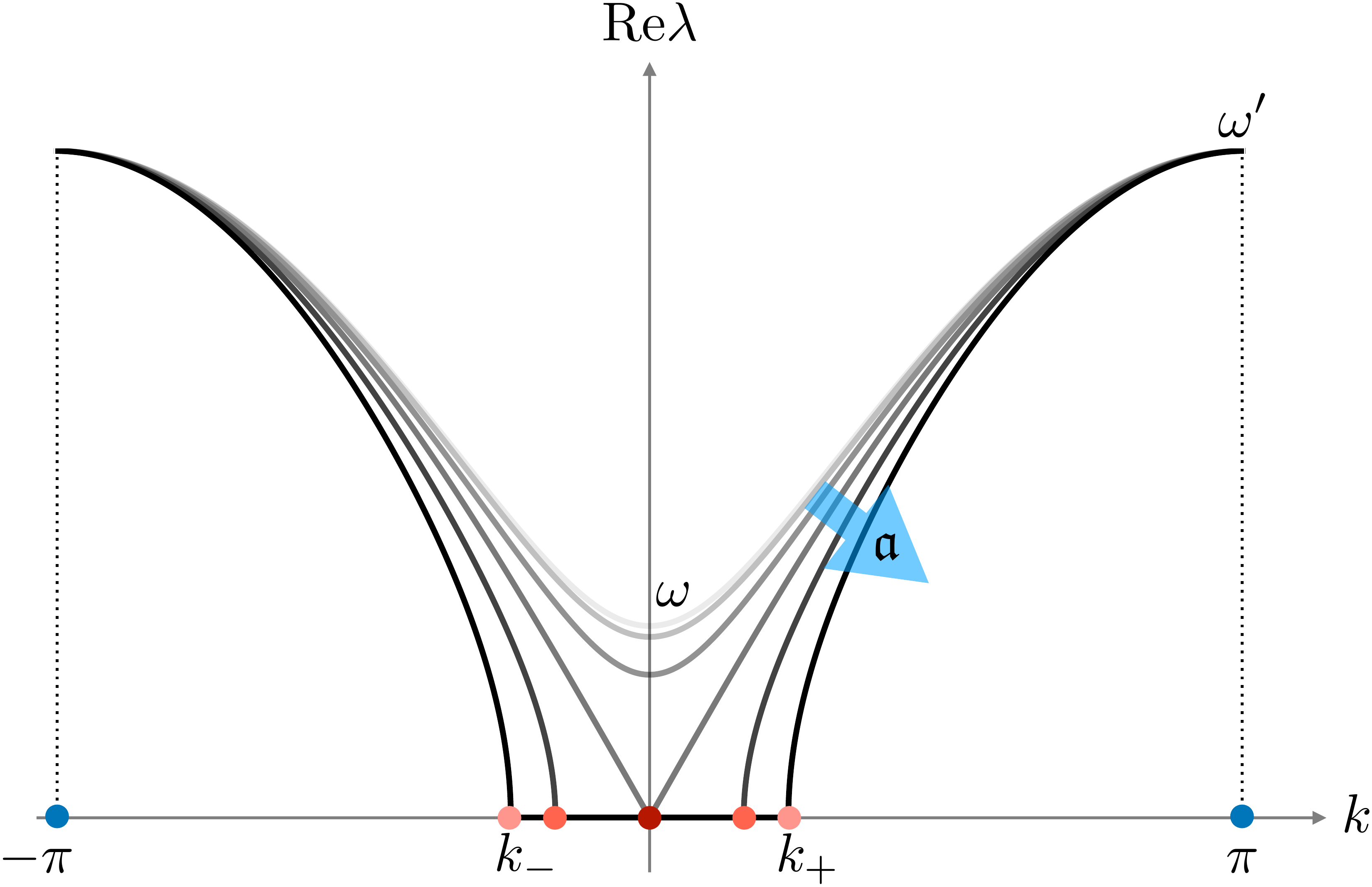}
\caption{The (positive real part of the) eigenfrequencies \rf{lAF}, with $k$ swept over the Brillouin zone. Here, we set $\omega/\omega'=0.3$ and increase $\fa$ from $0$ to $0.5$, in increments of $0.1$ (with the $0.3$ corresponding to the gapless dispersion). The EPs $k_\pm$, which solve Eq.~\rf{aEP}, are marked for $\fa=0.5$. Note how increasing the ordinary Gilbert damping $\fa$ closes the anisotropy gap $\omega$ in the intrinsic AF dispersion.}
\label{cz}
\end{figure}

Close to the EP, the frequency eigenvalues \rf{lAF} are given by
\eq{
\lambda_\pm\approx-i\fa\omega'\pm\sqrt{2(1-\fa\omega'/\omega_k)}\omega_k\,.
\label{aEP}}
This leads to a qualitatively similar behavior in the $k$-dependent eigenfrequency dispersions, as already discussed for the ferromagnetic case (cf. Fig.~\ref{bz}, along with the associated discussion). In particular, for $\fa<\omega/\omega'\equiv\sqrt{2K/\omega'}$, we obtain two dispersing modes with opposite circular polarizations and the same damping and frequency, for all wave numbers $k$. This is the ordinary AF dynamics. A stronger damping, $\fa>\sqrt{2K/\omega'}$, however, results in the synchronized (zero-frequency damped) dynamics within the two exceptional points $k_\pm$ (corresponding to $\omega_k/\omega'=\fa$), without any dispersion. Finally, a Weyl point with a closed gap is obtained at $\fa=\sqrt{2K/\omega'}$. In this critical case (corresponding to the AF resonance quality factor $\sim1$), an ordinary Gilbert damping closes the gap opened in the undamped AF dynamics by an easy-axis anisotropy, restoring the linear Goldstone-mode dispersion. We plot the positive real part of the dispersion \rf{lAF} in Fig.~\ref{cz}.

\section{Summary and outlook}
\label{so}

When embedded into a dissipative environment, coupled spin dynamics can yield exceptional points that have a drastic effect on their spectral properties. We focused  our discussion on ferromagnetic and antiferromagnetic dimers and spin chains. In the antiferromagnetic case, the EPs can be experimentally accessed by simply tuning the overall damping of the system. The ferromagnetic case requires a nonlocal dissipative coupling, which can be realized by internal magnetic spin pumping into itinerant degrees of freedom. In both cases, the EP emerges in the magnon band structure as a linearly dispersing Weyl point, which acts as a precursor to a flat magnon dispersion that extends over a finite range of momenta.

In the strongly-damped regimes, the EP singularities, therefore, separate the dispersing and nondispersing regions of the magnon band structure. These examples show how controlling dissipation can dramatically modify kinetic properties of magnons, such as a steplike change in their group velocities, in the vicinity of an EP. The resultant thermodynamic response and kinetic coefficients of the magnon gas, such as its spin conductivity and spin diffusion length \cite{cornelissenNATP15}, could then be invoked to exhibit this physics, in addition to coherent microwave probes.

Reducing the spins and crossing over to the quantum regime of the coupled dynamics, at low temperatures, it can be interesting to explore how the EPs evolve into the quantum limit of magnetic fluctuations. It may be intriguing, for example, to look for the associated features in the entanglement properties between individual spins \cite{zouCM19} or magnetic sublattices \cite{kamraPRB19}.

Braiding around the classical EPs in the parameter space taps into their Riemannian topological aspects, manifested, for example, in characteristic phase changes and non-Abelian braiding representation in the eigenvector evolution \cite{berryANY95}, along with the associated topological energy transfer \cite{xuNAT16}. When applying such braiding to the evolution of an open quantum spin system, one may be compelled to investigate the possibility of robust features inherited from the topology in the underlying classical counterpart, in regard, for example, to quantum gates and information processing tasks. In the future works, it could also be interesting to study the braiding of magnons in momentum space (or mixed parameter-momentum space), both in the classical and quantum regimes.

\begin{acknowledgments}
We thank Can-Ming Hu and Gerrit E. W. Bauer for drawing attention to this problem and Benedetta Flebus and Mostafa Tanhayi Ahari for helpful discussions. The work was supported by NSF under Grant No. DMR-1742928.
\end{acknowledgments}

%

\end{document}